\documentclass[preprint2]{aastex}

\shorttitle{Supergranules as Probes of the Sun's Meridional Circulation}
\shortauthors{Hathaway}

\begin{document}


\title{Supergranules as Probes of the Sun's Meridional Circulation}
\author{David H. Hathaway}
\affil{NASA Marshall Space Flight Center, Huntsville, AL 35812 USA}
\email{david.hathaway@nasa.gov}

\begin{abstract}
Recent analysis revealed that supergranules (convection cells seen at the Sun's surface) are advected by the zonal flows at depths equal to the widths of the cells themselves.
Here we probe the structure of the meridional circulation by cross-correlating maps of the Doppler velocity signal using a series of successively longer time lags between maps.
We find that the poleward meridional flow decreases in amplitude with time lag and reverses direction to become an equatorward return flow at time lags $> 24$ hours.
These cross-correlation results are dominated by larger and deeper cells at longer time lags. 
(The smaller cells have shorter lifetimes and do not contribute to the correlated signal at longer time lags.)
We determine the characteristic cell size associated with each time lag by 
comparing the equatorial zonal flows measured at different time lags with the zonal flows associated with different cell sizes from a Fourier analysis.
This association gives a characteristic cell size of $\sim 50$ Mm at a 24 hour time lag.
This indicates that the poleward meridional flow returns equatorward at depths $> 50$ Mm -- just below the base of the surface shear layer.
A substantial and highly significant equatorward flow (4.6 $\pm$ 0.4 m s$^{-1}$) is found at a time lag of 28 hours corresponding to a depth of $\sim 70$ Mm.
This represents one of the first positive detections of the Sun's meridional return flow and illustrates the power of using supergranules to probe the Sun's internal dynamics.

\end{abstract}


\section{INTRODUCTION}

Supergranules are cellular flows observed in the Sun's photosphere as a surface filling pattern of largely horizontal flows \citep{RieutordRincon10}.
Typical supergranules have diameters of $\sim 30$ Mm and lifetimes of $\sim 1$ day -- much larger and longer lived than granules (diameters of $\sim$ 1 Mm and lifetimes of $\sim$ 5 minutes).

Since their discovery by \cite{Hart54} and their initial characterization by \cite{Leighton_etal62}, supergranules have been recognized as the principal organizer of the Sun's chromospheric/magnetic network and as key to the transport of magnetic flux in the photosphere \citep{Leighton64, DeVore_etal84}.

The rotation rate of the supergranules has been measured by a number of investigators with interesting results.
\cite{Duvall80} cross-correlated the Doppler velocity pattern from equatorial spectral scans obtained over several days and found that the pattern rotates about 3\% faster than the photospheric plasma.
Faster rates are found for the 24-hr time lags from day-to-day than for the 8-hr time lags from the beginning to end of an observing day.
\cite{BeckSchou00} measured the rotation of the Doppler velocity pattern using a Fourier technique and found that the larger cells rotate more rapidly than the smaller cells but with apparent rotation rates that exceed the peak internal rotation rate.
This ``superrotation'' was attributed to line-of-sight projection effects on the Doppler velocity pattern by \cite{Hathaway_etal06}.
In fact, \cite{Schou03} had earlier found rotation consistent with the internal plasma flow profile when this line-of-sight projection was removed.

These observations are consistent with the original conclusion by \cite{Duvall80} that larger cells dominate the longer time lags and that these larger cells are more deeply anchored in a surface shear layer in which the rotation rate increases with depth.
While it should be noted that some characteristics of the motions of the supergranules have been attributed to wave-like properties \citep{Gizon_etal03, Schou03}, it is clear that the cellular structures are advected largely by the zonal flows in the surface shear layer.

The existence of this shear layer was suggested by \cite{FoukalJokipii75} as a consequence of the conservation of angular momentum by convective elements moving inward and outward in the near surface layers.
Helioseismology \citep{Thompson_etal96, Schou_etal98} now indicates that the shear layer extends to a depth of $\sim50$ Mm.

Recently \cite{Hathaway12} found a one-to-one correspondence between the rotation rate of supergranules with increasing size and the rotation rate in this surface shear layer with increasing depth.
Results, using a Fourier technique like that of \cite{BeckSchou00} with the line-of-sight projection effects removed using the method of \cite{Schou03}, indicate that supergranules are advected by the flows in the shear layer at depths equal to their widths.
We expect the advection of the supergranules to be dominated by flows near their bases since the plasma density increases rapidly with depth.
However, this does imply that the cells must extend even deeper than this anchoring level. 

Note that while cells that are as deep as they are wide are typical in numerical simulations of solar convection \citep{Stein_etal11}, results from local helioseismology have suggested flattened supergranules -- but with conflicting results.
\cite{Duvall98} found that the outflows in the supergranules reversed at a depth of 5 Mm and then disappeared by 8 Mm depth.
\cite{ZhaoKosovichev03} found reversal at a depth of 8 Mm and disappearance at 15 Mm.
\cite{Svanda_etal09} found disappearance at an even greater depth -- 25 Mm.

Here we measure the meridional motion of the pattern of supergranules by analyzing the same data used in \cite{BeckSchou00}, \cite{Schou03}, and \cite{Hathaway12}.

The Sun's meridional flow is extremely weak when compared to the other photospheric flows.
Both the strength and the direction of the meridional flow were uncertain until data and data analysis techniques were sufficiently improved \citep{Komm_etal93, Hathaway_etal96, Giles_etal97, Ulrich10}.
The meridional flow is now known to be poleward in each hemisphere with a peak velocity of about 10-20 m s$^{-1}$ while the zonal flows (differential rotation) are an order of magnitude stronger and the convective flows (granules and supergranules) are yet another order of magnitude stronger.

The Sun's meridional flow plays a critical role in the transport of magnetic flux \citep{DeVore_etal84, Wang_etal09} and in some dynamo models for the Sun's magnetic cycle \citep{DikpatiCharbonneau99, NandyChoudhuri02}.
Although the nonaxisymmetric convective motions transport magnetic elements at much higher speeds to the cell boundaries, those motions are in random directions and give diffusion away from flux concentrations.
The net poleward transport has a larger contribution from the slow, but direct, meridional flow.
This poleward transport is directly responsible for the reversal of the Sun's polar magnetic fields every solar cycle and for producing the polar fields that determine the strength of the following solar cycle \citep{Schatten_etal78, Svalgaard_etal05}.

Fourier techniques (which do give explicit information on the cell sizes) are, unfortunately, not available for measuring the meridional motion of the supergranule Doppler velocity pattern.
Here we use cross-correlations with time lags that vary from 1 hour to 28 hours.
The displacement of the peak in the cross-correlation gives both zonal and meridional velocities which vary systematically with time lag.
We determine the characteristic cell size for each time lag by comparing the zonal velocities with those obtained with the Fourier technique \citep{Hathaway12}.
This then gives the meridional flow at a series of anchoring depths corresponding to the series of time lags.

\section{DATA PREPARATION}

The data consist of $1024^2$ pixel images of the line-of-sight velocity determined from the Doppler shift of a spectral line due to the trace element nickel in the solar atmosphere
by the Michelson Doppler Imager \citep[MDI]{Scherrer_etal95} on the ESA/NASA {\it Solar and Heliospheric Observatory (SOHO)}.

The images are acquired at a 1 minute cadence and cover the full visible disk of the Sun.
We average the data over 31 minutes with a Gaussian weighting function which filters out variations on time scales less than about 16 minutes, and sample that data at 15 minute intervals.
We then map these temporally filtered images onto a $1024^2$ grid in heliographic latitude from pole to pole and in longitude $\pm90\degr$ from the central meridian (Figure 1). This mapping accounts for the position angle of the Sun's rotation axis relative to the imaging CCD and the tilt angle of the Sun's rotation axis toward or away from the spacecraft. Both of these angles include modifications in line with the most recent determinations of the orientation of the Sun's rotation axis \citep{BeckGiles05, HathawayRightmire10}. We analyze data obtained during two 60 day periods of continuous coverage: one in 1996 from May 24 to July 22; the other in 1997 from April 14 to June 17.

Each filtered image is analyzed using the methods described by \cite{Hathaway92} to determine
and remove the stationary signals due to differential rotation, convective blue shift, and the meridional flow -- as well as the time varying signal due to the motion of the spacecraft.
Following this, an average image from all of the images in each 60 day period was subtracted from each image to remove any instrumental artifacts.
This leaves behind the non-axisymmetric and time-varying signal due to the convective flows.

\begin{figure}[htb]
\centerline{\includegraphics[width=\columnwidth]{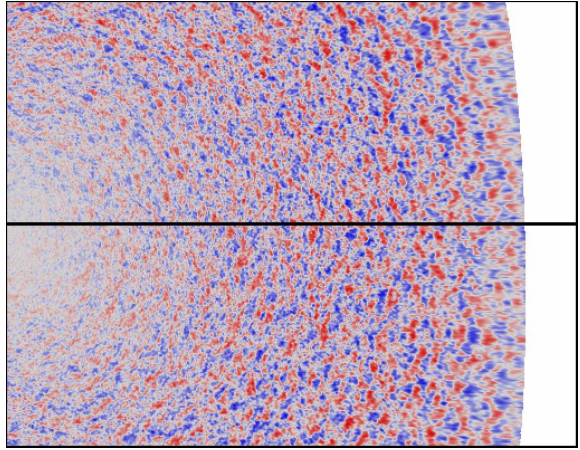}}
\caption{
Heliographic map detail of the line-of-sight (Doppler) velocity from SOHO/MDI.
The map detail extends $90\degr$ in longitude from the central meridian on the left and about $35\degr$ in latitude from the equator (the thick horizontal line).
The mottled pattern is the Doppler signal (blue for blue shifts and red for red shifts) due to the supergranules.}
\end{figure}

\section{CROSS-CORRELATION ANALYSIS}

We determine the zonal and meridional velocity of the supergranule pattern by cross-correlating strips of data 11 pixels high in latitude by 601 pixels long in longitude.
The strips are centered on each of the 860 latitude positions from $75\degr$ south to $75\degr$ north of the equator with overlap from adjacent strips.
These data strips are cross-correlated with corresponding strips from images obtained at time lags of 1, 2, 4, 8, 16, 24, and 28 hours.

This process is done both forward and backward in time.
A data strip centered on the central meridian is correlated with strips offset to the left at later times and then data centered on the central meridian at later times is correlated with
strips offset to the right in the initial data map.
This minimizes any systematic errors.

The shift of the Doppler pattern is determined to the nearest pixel by finding the displacement of the strip in the later Doppler map that gives the maximum correlation.
This pixel-wise shift is then refined by fitting parabolas through the correlation amplitudes at the displacement that gives the maximum correlation and those at plus and minus one pixel in each direction.
The maxima of these parabolas then give the Doppler pattern shift to a fraction of a pixel.
The shifts in longitude give the apparent zonal velocity while the shifts in latitude give the meridional velocity.
The 60 days of data in each data set give some 1400 independent measurements of these velocities at each of the 860 latitude positions for each time lag.

The meridional flow measurements for time lags from 1 to 24 hours are shown in Figure 2.
Here the data are smoothed in latitude with a tapered Gaussian having a FWHM of $10\degr$.
(A tapered Gaussian is a Gaussian with a quadratic function subtracted so that the smoothing filter and its first derivative go to zero at the endpoints.) 
The flow is poleward in each hemisphere from the equator up to latitudes of at least $60\degr$.
The speed of the flow decreases as the time lag increases and the poleward flow virtually vanishes when measured with the 24 hour time lags.
The slow down for time lags from 2 to 16 hours was previously noted by \cite{Hathaway_etal10}.
The lack of a poleward meridional flow for 24 hour time lags was previously noted (and referred to as ``anomalous'') by \cite{Gizon_etal03}.

\begin{figure}[ht]
\centerline{\includegraphics[width=0.7\columnwidth]{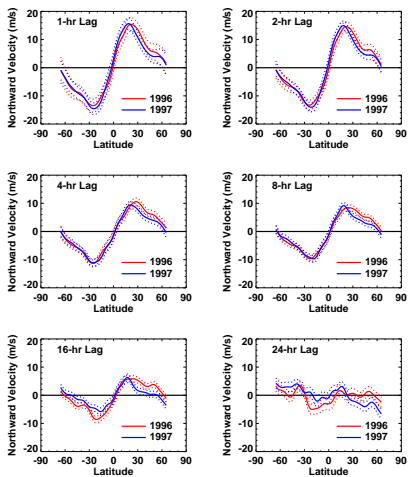}}
\caption{
Meridional flow profiles for time lags from 1 to 24 hours.
The measurements from the 1996 data are shown with the red line while the measurements from the 1997 data are shown with the blue line.
The dotted lines indicate $2\sigma$ error limits for each year.
The meridional flow is poleward up to latitudes of at least $60\degr$.
The amplitude of this poleward meridional flow decreases as the time lag increases and disappears at 24 hour time lags.}
\end{figure}

Measurements with 28 hour time lags are shown in Figure 3.
Again, the data are smoothed in latitude with a tapered Gaussian having a FWHM of $10\degr$.
With this longer time lag the measured meridional flow is {\it equatorward} at virtually all latitudes with a peak of $\sim4$ m s$^{-1}$.
The $2\sigma$ variations on the measurements at each latitude are $\sim2$ m s$^{-1}$ making nearly all of the individual latitude measurements statistically significant.
Fitting the measurements to a 4th-order polynomial in $\sin \theta$, shown by the thick black line in Figure 3, gives a peak velocity of 4.6 $\pm$ 0.4 m s$^{-1}$ with $2\sigma$ errors.
As such this represents the first positive measurement and significant ($>10\sigma$) detection of the Sun's equatorward return flow. 

\begin{figure}[ht]
\centerline{\includegraphics[width=\columnwidth]{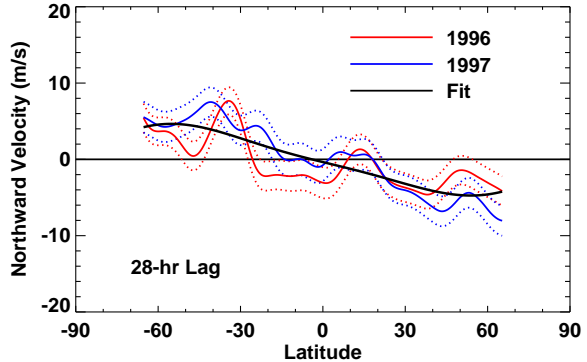}}
\caption{
Meridional flow profiles for time lags of 28 hours.
The measurements from the 1996 data are shown with the red line while the measurements from the 1997 data are shown with the blue line.
The dotted lines indicate $2\sigma$ error limits for each year.
The 4th-order fit to the combined data is shown with the thick, black line.}
\end{figure}

\section{DEPTH DETERMINATIONS}

\cite{Duvall80} suggested that at different time lags the cells that survive are advected by flows with different speed (and direction) depending upon the depth to which those cells extend. The different time lags thus probe the depth dependence of the meridional flow.
\cite{Hathaway12} found that supergranules of a given cell size (given by their wavenumber in a Fourier analysis) are advected by the zonal flows at depths equal to their widths.
This was determined by comparing their zonal velocities from the Fourier analysis with zonal velocities in the surface shear layer determined with global helioseismology by \cite{Schou_etal98}.

Small cells have short lifetimes.
Granules with typical diameters of $\sim1$ Mm have lifetimes of $\sim5$ minutes \citep{Title_etal89}.
Supergranules with typical diameters of $\sim30$ Mm have lifetimes of $\sim1$ day \citep{SimonLeighton64, WangZirin89, Hirzberger_etal08}.
Cells of intermediate (mesogranular) size with diameters 5-10 Mm have intermediate lifetimes of $\sim2$ hours \citep{November_etal81}.
As the time lag used in the cross-correlation increases, the cells with lifetimes shorter than the time lag present an uncorrelated random pattern that does not influence the determination of the pattern motion. Longer time lags are associated with larger cells which extend deeper into the Sun.
This assertion is supported by observations of the monotonic decline in the cross-correlations for granules \citep{Title_etal89}, the monotonic decline in the cross-correlations for the supergranule Doppler pattern \citep{Hathaway_etal10},and the increase in lifetimes for larger supergranule cells determined from local helioseismology \citep{Hirzberger_etal08}.

We can determine the characteristic cell size at each time lag by matching the zonal velocities from the cross-correlations with those from the Fourier method.
This Fourier method is described in detail by \cite{BeckSchou00} and \cite{Hathaway12}.
Lines of mapped Doppler data at equatorial latitudes are Fourier transformed in longitude to give spectral coefficients.
Spectral coefficients from 6 10-day series of Doppler maps obtained at 15-minute cadence are then Fourier transformed in time.
The location of the peak power gives the zonal (longitudinal) velocity for each spatial wavelength or cell size.

The line-of-sight projection for the Doppler signal influences the measured zonal flow (but, as shown by \cite{Hathaway_etal10}, not the meridional flow).
When this projection is divided-out (as can only be done at the equator) the zonal flow as a function of wavelength from the Fourier analysis matches the zonal flow as a function of depth from global helioseismology.
Figure 4 shows zonal velocities relative to the Carrington Rotation frame of reference for supergranules of varying wavelength along with the zonal velocity as a function of depth from global helioseismology \citep{Schou_etal98}.

\begin{figure}[ht]
\centerline{\includegraphics[width=\columnwidth]{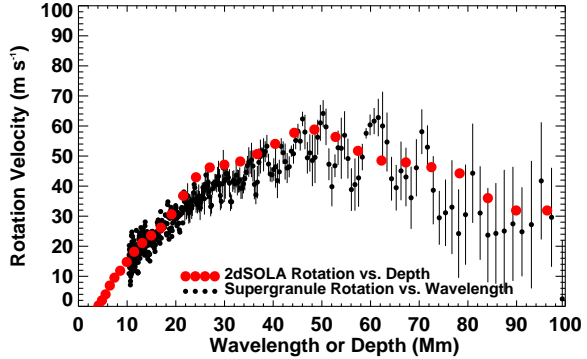}}
\caption{
Equatorial zonal velocities as functions of cell wavelength and depth within the Sun.
Zonal velocities as functions of wavelength from the 1996 MDI data in which the line-of-sight projection is divided out are shown with the small black circles.
Zonal velocities as functions of depth from global helioseismology \citep{Schou_etal98} are shown with the large red circles.
The match between these measurements at virtually all wavelengths and depths indicates that supergranules are advected by the flows at depths equal to their wavelengths.}
\end{figure}

The zonal velocities for the raw data (without the line-of-sight projection removed) are shown in Figure 5.
Comparing these zonal velocities with those obtained at the equator (also with the raw data) by the cross-correlation analysis gives a correspondence between time lag and characteristic wavelength of the cells that dominate at that time lag.
This correspondence is given in Table 1.

\begin{figure}[ht]
\centerline{\includegraphics[width=\columnwidth]{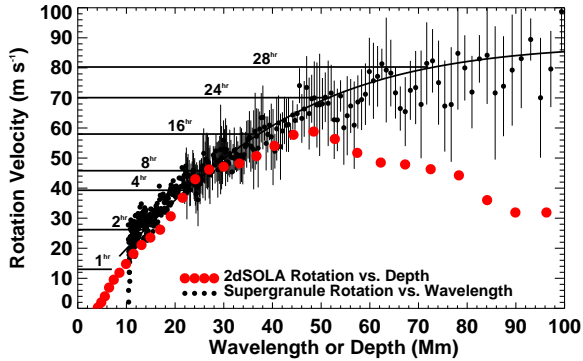}}
\caption{
Equatorial zonal velocities as functions of cell wavelength and depth within the Sun.
Zonal velocities as functions of wavelength from the 1996 MDI data are shown with the small black circles.
Zonal velocities as functions of depth from helioseismology \citep{Schou_etal98} are shown with the large red circles.
The zonal velocities at the equator from the different time lags are indicated along the vertical axis with lines indicating where they match the zonal velocities from the Fourier analysis.}
\end{figure}

\begin{table}[htb]
	\centering
		\begin{tabular}{ccc}
			Time Lag & Velocity & Wavelength \\ \hline
			1 hour & $12\pm1$ m s$^{-1}$ & $7\pm1$ Mm \\
			2 hours & $24\pm1$ m s$^{-1}$ & $14\pm1$ Mm \\
			4 hours & $40\pm1$ m s$^{-1}$ & $22\pm1$ Mm \\
			8 hours & $49\pm2$ m s$^{-1}$ & $26\pm2$ Mm \\
			16 hours & $59\pm2$ m s$^{-1}$ & $37\pm2$ Mm \\
			24 hours & $71\pm3$ m s$^{-1}$ & $50\pm5$ Mm \\
			28 hours & $80\pm3$ m s$^{-1}$ & $70\pm5$ Mm \\
		\end{tabular}
		\caption{Correspondence between time lag, equatorial zonal velocity, and cell wavelength from Figure 5.}
\end{table}

\section{CONCLUSIONS}

We conclude that the Sun's poleward meridional flow is confined to it's surface shear layer (from the photosphere to a depth of $\sim50$ Mm).
We draw this conclusion from the fact that the poleward flow weakens and then disappears as the cross-correlation time lag increases from 1 to 24 hours while the measured rotation rates indicate that the cross-correlated patterns are dominated by cells with sizes, and therefor depths, that increase from $\sim7$ Mm to $\sim50$ Mm.
A positive detection of the equatorward return flow is found with time lags of 28 hours corresponding to depths of $\sim70$ Mm.

The fact that the return flow signal was significantly detected at a longer time lag in a deeper layer also indicates that the lack of a poleward flow at the 24 hour time lag is due to an actual lack of poleward flow rather than an insensitivity to the signal.

Estimating the poleward mass flux at $30\degr$ latitude within the surface shear layer using the measured flow speeds and the run of plasma density with depth, indicates that a balancing equatorward mass flux at a flow speed of $\sim4$ m s$^{-1}$ should only extend another 20-30 Mm below the base of this surface shear layer.

This shallow return flow is somewhat surprising given earlier results from helioseismology.
\cite{BraunFan98} used frequency shifts between northward and southward propagating waves and concluded that the meridional flow was poleward to a depth of at least 100 Mm.
\cite{Giles00} used time-distance helioseismology with a mass flux constraint and concluded that the return flow was below a depth of about 140 Mm.
However, \cite{DuvallHanasoge09} have found that these measurements are plagued by systematic errors leading \cite{GoughHindman10} to state that, with helioseismology, the meridional flow below a depth of 30 Mm remains uncertain.
 
Recently \cite{Zhao_etal12A} have characterized these systematic signals, removed them from the data, and retrieved a meridional flow profile in the near surface layers that agrees well with measurements obtained from magnetic feature motions \citep{HathawayRightmire10} and direct Doppler measurements \citep{Ulrich10}.
\cite{Zhao_etal12B} have now reported that they find evidence for the meridional return flow at depths much shallower than the tachocline -- supporting the conclusions reached here.

A shallow return flow is also supported by \cite{Sivaraman_etal10} who measured the meridional motions of sunspot umbrae.
They compared the rotation rates of sunspot umbrae with rotation rates derived from helioseismology to find anchoring depths for umbrae of different sizes (similar to what was done here with supergranules) and found that small sunspots
with umbral areas $< 5 \mu$Hem (a $\mu$Hem is $10^{-6}$ the area of a solar hemisphere) are anchored at depths of $\sim 90$ Mm and are moving equatorward at speeds of $\sim 7$ m s$^{-1}$.
Slightly deeper anchoring depths and slower equatorward velocities were found for larger sunspots.

\section{FURTHER IMPLICATIONS}

The shallow depth of the meridional return flow has implications for our understanding of both solar convection zone dynamics and solar dynamo theory.

The axisymmetric meridional flow is maintained by a balance between the combined effects of three driving forces (latitudinal pressure gradients, Reynolds stresses resulting from correlations between radial and meridional motions in the cellular flows, and the Coriolis force acting on the axisymmetric zonal flow) and the viscous diffusion due to the cellular flows.
Simulations of the effects of the Sun's rotation on supergranules in plane-parallel geometry \citep{Hathaway82} indicate that the Reynolds stresses produced by the supergranules themselves can maintain both the zonal shear (slower rotation at the surface and faster rotation below) and a meridional circulation with poleward flow near the surface and equatorward flow below.
The meridional Reynolds stresses (correlated poleward and upward motions in the cellular flows) are reinforced by the Coriolis force on the axisymmetric zonal flow in these simulations.
(Latitudinal pressure gradients cannot be maintained or included due to the periodic boundaries in both horizontal directions.)

The meridional circulations produced by larger convection cells in rotating spherical shells extending over several density scale heights (but thus far not including the surface shear layer) are usually highly structured in latitude and radius and are highly variable in time \citep{Miesch_etal00}.
More recent spherical shell simulations \citep{Miesch_etal08} impose a latitudinal entropy (temperature) gradient at the base of the convection zone to give rotation profiles and meridional circulations more in line with observations but still produce highly time dependent meridional flows.
The source of this structure and variability can be attributed to the small number ($\sim 100$) of convection cells that populate the simulated global volume.
A meridional circulation driven in the surface shear layer by ($\sim 10000$) supergranules (the convection cells that populate the surface shear layer) is far more likely to be less structured and variable -- as is observed.

The shallow return flow we find here has implications for flux transport dynamos \citep{DikpatiCharbonneau99, NandyChoudhuri02} -- it violates a key assumption.
These dynamo models assume that the return flow begins at a depth of 100 Mm and reaches a maximum equatorward speed of just 1-2 m s$^{-1}$ at the base of the convection zone.
This is a critical assumption in that the equatorward drift of the sunspot zones is directly produced in these dynamo models by this slow equatorward flow at the base of the convection zone.

The results presented here support the idea of using supergranules to probe the dynamics of the solar convection zone.
This method can complement the probing done with acoustic waves by the methods of helioseismology and provide valuable information on the dynamics of solar and stellar convection zones.
The method used here, finding the peak in the cross-correlated Doppler velocity pattern for data pairs with different time lags, is limited to time lags less than $\sim32$ hours.
At longer time lags the correlation amplitudes from individual data pairs are similar to the noise level from the cross-correlations.
However, averaging the cross-correlations over as few as 24 hourly measurements could extend the results to longer time lags and deeper layers.
Local Correlation Tracking using smaller patches might also yield measurements of the large  scale nonaxisymmetric flows -- Giant Cells.

Finally, it should be noted that, while these results indicate an equatorward return flow at a fairly shallow depth, they do not preclude the possible existence of additional cells of poleward and equatorward flow at even greater depths.

\acknowledgements
The author would like to thank NASA for its support of this research through grants
from the Heliophysics Causes and Consequences of the Minimum of Solar Cycle 23/24 Program and the Living With a Star Program to NASA Marshall Space Flight Center. He is indebted to Ron Moore and Lisa (Rightmire) Upton and an anomymous referee whose comments greatly improved the manuscript and to John Beck who produced the temporally filtered data from the original MDI Doppler data. He would also like to thank the American taxpayers who support scientific research in general and this research in particular. SOHO, is a project of international cooperation between ESA and NASA.

\end{document}